\documentclass[aps,prd,showpacs,superscriptaddress,floatfix,cite,twocolumn,nofootinbib]{revtex4-2}
\usepackage{amsmath,amssymb,bm,color,fancyhdr,hyperref,lastpage,units,soul}
\usepackage{graphicx}

\usepackage[latin1,utf8]{inputenc}
\usepackage[english]{babel}
\usepackage{makecell}
\usepackage{float}

\newcommand{\mY}{\mu_Y}
\newcommand{\hmY}{\hat{\mu}_Y}
\newcommand{\mI}{\mu_I}
\newcommand{\hmI}{\hat{\mu}_I}

\newcommand{\mB}{\mu_B}
\newcommand{\hmB}{\hat{\mu}_B}
\newcommand{\dP}{\Delta P}
\newcommand{\mT}{\left(\frac{\mY}{T}\right)}
\newcommand{\Ns}{N_\sigma}
\newcommand{\Nt}{N_\tau}
\newcommand{\Nrv}{{N_\text{rv}}}
\newcommand{\Ob}{\mathcal{O}}

\newcommand{\A}{\mathcal{A}}
\newcommand{\C}{\mathcal{C}}

\newcommand{\K}{\mathcal{K}}
\renewcommand{\L}{\mathcal{L}}
\newcommand{\M}{\mathcal{M}}
\newcommand{\N}{\mathcal{N}}
\renewcommand{\O}{\mathcal{O}}
\newcommand{\W}{\mathcal{W}}

\newcommand{\Z}{\mathcal{Z}}

\newcommand{\Tpc}{T_{pc}}

\def\lsim{\raise0.3ex\hbox{$<$\kern-0.75em\raise-1.1ex\hbox{$\sim$}}}
\def\gsim{\raise0.3ex\hbox{$>$\kern-0.75em\raise-1.1ex\hbox{$\sim$}}}

\renewcommand{\Re}{\text{Re}}
\renewcommand{\Im}{\text{Im}}

\begin{document}
\title{QCD equation of state at finite chemical potential from unbiased \\exponential resummation of the lattice QCD Taylor series}

\author{Sabarnya Mitra}
\affiliation{Centre for High Energy Physics, Indian Institute of Science, Bengaluru 560012, India.}

\author{Prasad Hegde}
\affiliation{Centre for High Energy Physics, Indian Institute of Science, Bengaluru 560012, India.}
\email[]{prasadhegde@iisc.ac.in}

\begin{abstract}
Exponential resummation of the QCD finite-density Taylor series has been recently introduced as an alternative way of resumming the finite-density lattice QCD Taylor series. Unfortunately the usual exponential resummation formula suffers from stochastic bias which must be subtracted before identifying genuine higher-order contributions. In this paper, we present a new way of subtracting the stochastic bias at the level of each individual gauge configuration, up to a certain order of either the Taylor series or the cumulant expansion, by modifying the argument of the exponential. Retaining the exponential form of the resummation allows us to also calculate the phase factor of the fermion determinant on each gauge configuration. We present our results for the excess pressure, number density, and the average phase factor and show that the new results contain less stochastic bias and are in better agreement with the QCD Taylor series compared to the previous exponential resummation.
\end{abstract}

\maketitle

\section{Introduction}
\label{sec:introduction}
The phase diagram of strongly interacting matter as a function of the temperature $T$ and baryochemical potential $\mB$ is of interest to theorists and experimentalists alike~\cite{STAR:2010vob,Guenther:2020jwe}. Since the system is non-perturbative except at very large temperatures and chemical potentials, a reliable non-perturbative approach is required for its study. At $\mu_B=0$, such an approach is provided by lattice QCD. In recent years, lattice calculations have provided increasingly precise determinations of several properties of the quark-gluon plasma~\cite{HotQCD:2014kol,Bazavov:2017dus,HotQCD:2018pds,Bollweg:2021vqf,Dimopoulos:2021vrk,Bollweg:2022rps}. Unfortunately however, lattice QCD breaks down at $\mB\ne0$ due to the well-known sign problem~\cite{deForcrand:2009zkb, Aarts:2014fsa, Aarts:2015kea, Nagata:2021bru}. Despite recent progress~\cite{Aarts:2009yj, Cristoforetti:2012su, Aarts:2013uxa, Sexty:2013ica, Fukuma:2019uot, Giordano:2020roi}, currently the two most successful approaches in the QCD case are analytical continuation from imaginary to real $\mu_B$~\cite{Borsanyi:2018grb,Ratti:2018ksb} and Taylor expansion of the QCD partition function in the chemical potential $\mu_B$~\cite{Bazavov:2017dus,Bollweg:2021vqf}. Despite their successes however, both methods need to be supplemented in order to obtain reliable results beyond $\hmB \equiv\mB/T\simeq1$-$2$~e.g.~by combining the results at imaginary $\mB$ with an alternative expansion scheme~\cite{Borsanyi:2021sxv} or by resumming the QCD Taylor series through the use of Pad\'e resummation~\cite{Datta:2016ukp,Giordano:2019slo,Dimopoulos:2021vrk,Bollweg:2022rps}.

An alternative way of resumming the QCD Taylor series was recently proposed in Ref.~\cite{Mondal:2021jxk}. The calculation of the Taylor coefficients requires the $n$th $\hmB$ derivative $D^B_n$ of $\ln\det\M$, where $\hmB\equiv\mB/T$ and $\det\M$ is the fermion matrix determinant. The contribution of $D_n^B$ to all orders of the Taylor series can be shown to be $\exp\,(D_n^B\hmB^n/n!)$. Resumming the first $N$ derivatives in this way leads to an improved estimate for the QCD Equation of State (QEOS) which is equal to the $N$th order Taylor estimate plus all the higher order contributions coming from $D^B_1,\dots,D_N^B$. It can be shown that the resummed QEOS indeed captures some of the contributions coming from the higher-order Taylor coefficients~\cite{Mondal:2021jxk}. Furthermore, since the odd (even) $D^B_n$ are purely imaginary (real), the resummation procedure yields an estimate for the complex phase factor of the fermion determinant. The ensemble-averaged phase factor $\big\langle e^{i\Theta(T,\mu_B)}\big\rangle$ goes to zero as $\mB$ is increased due to which the calculation of the resummed QEOS breaks down. This breakdown is physical and can be related to the presence of poles or branch cut singularities of the QCD partition function in the complex $\mB$ plane. The resummation approach also makes it possible to calculate these singularities directly. Some of these advantages have been previously demonstrated through analytical calculations in a low-energy model of QCD~\cite{Mukherjee:2021tyg}.

Despite its advantages, one drawback of exponential resummation in the lattice QCD case is the presence of stochastic bias in the calculation of the exponential factor. Given $N$ independent random estimates $W_1,\dots,W_N$ of an observable $\W$, the unbiased estimate of $\W^n$ is given by
\begin{equation}
\text{UE}\left[\W^n\right] = \sum_{i_1\ne i_2\ne\dots\ne i_n} \frac{W_{i_1} \cdots W_{i_n}}{N(N-1)\cdots(N-n+1)}.
\end{equation}
That is, an unbiased estimate is formed by averaging over products of independent estimates. The contribution of products of the same estimate is the stochastic bias, as in the biased estimate of $\W^n$ e.g.
\begin{equation}
\text{BE}\left[\W^n\right] = \left[\frac{1}{N}\sum_{i=1}^NW_i\right]^n.
\end{equation}
\par Although stochastic bias vanishes in the limit $N\to\infty$, for finite $N$ it can be comparable to the true value and hence lead to a wrong estimate in some cases. We shall see in Sec.~\ref{sec:formalism} that the usual formula for the exponential factor in exponential resummation contains stochastic bias. Subtracting this bias therefore becomes necessary, especially at higher orders and for large values of $\hmB$.

Unlike exponential resummation, stochastic bias is not a problem in the Taylor coefficient calculations because there exist efficient formulas for evaluating the unbiased product of $n$ operators in $\O(N)$, rather than $\O(N^n)$, time. Therefore one way to avoid stochastic bias, while still going beyond the Taylor series approach, is to replace exponential resummation by a finite order cumulant expansion~\cite{Mitra:2022vtf}. This approach corrects for stochastic bias but at the expense of all-orders resummation~\footnote{It is also possible to avoid stochastic bias by calculating the $D_n^B$ exactly~\cite{Borsanyi:2022soo}. However straightforward diagonalization is expensive, even with the reduced matrix formalism, and one is therefore constrained to work with lattices having a smaller aspect ratio than the lattices considered here.}. Additionally, a knowledge of the phase factor is also lost. Lastly, knowledge of the analytic structure of the QCD partition function is also lost since the cumulant expansion is a finite polynomial and is hence analytic over the entire complex $\mB$ plane.

At present, we know of no way of obtaining a fully unbiased estimate of a transcendental function such as the exponential. Nevertheless, in this paper we will present a way of subtracting the stochastic bias to a finite order of either the Taylor or the cumulant expansion while also simultaneously retaining the exponential form of the resummation. The formalism presented here thus manages to preserve all-orders resummation. Moreover, depending upon the order of the calculation and the value of $\hmB$, it may be sufficient if the bias is eliminated up to some finite order $N$. In that case, our formalism yields results that are close to fully unbiased resummation.

Our paper is organized as follows: In Sec.~\ref{sec:formalism}, we will outline the construction of the unbiased exponential. We will begin by discussing Taylor expansion, simple (biased) exponential resummation and the cumulant expansion. We will then show how to modify the argument of the exponential so that the stochastic bias is subtracted either to order $N$ of the Taylor series expansion or to some order $M$ of the cumulant expansion. The corresponding formulas are Eqs.~\eqref{eq:unbiased_resummed-1}, \eqref{eq:Cn} and Eqs.~\eqref{eq:unbiased_resummed-2}, \eqref{eq:Lm} respectively. However, we defer a proof of the unbiasedness of the former to Appendix~\ref{app:proof}. After presenting the formalism, in Sec.~\ref{sec:results} we will present results for the excess pressure and number density for both finite isospin as well as baryochemical potential up to fourth order in the Taylor, biased resummation and unbiased resummation approaches. We will also present results for the average phase factor calculated using biased as well as unbiased resummation. Finally, in Sec.~\ref{sec:discussion_and_outlook}, we will summarize our results and conclusions.

\section{Unbiased Exponential Resummation}
\label{sec:formalism}
Consider lattice QCD with $2+1$ flavors of rooted staggered quarks defined on an $\Ns^3\times\Nt$ lattice. The partition function $\Z(T,\mY)$ at temperature $T$ and finite chemical potential $\mY$ is given by 
\begin{equation}
    \Z(T,\mY) = \int \mathcal{D}U e^{-S_G(T)} \,\det \M(T,\mY),
    \label{eq:partition_function}
\end{equation}
where $S_G(T)$ is the gauge action. The finite baryochemical potential $\mB$ case corresponds to $Y=B$ whereas the finite isospin chemical potential case corresponds to $Y=I$. The fermion determinant $\det\M(T,\mY)$ is given by
\begin{equation}
    \det\M(T,\mY) = \prod_{f=u,d,s}\big[\det \M_f(m_f,T,\mu_f)\big]^{1/4},
    \label{eq:fermion_determinant}
\end{equation}
with $m_u=m_d$ and $\mu_u$, $\mu_d$ and $\mu_s$ chosen appropriately according to $Y=B,I$\footnote{$\mu_u=\mu_d=\mu_s=3\,\mu_B$ for $Y=B$. For $Y=I$, $\mu_u=-\mu_d=\mu_I$ and $\mu_s=0$.}. The excess pressure $\dP(T,\mY) \equiv P(T,\mY)-P(T,0)$ is given by
\begin{equation}
    \frac{\dP(T,\mY)}{T^4} = \frac{1}{VT^3} \, \ln \left[\frac{\Z(T,\mY)}{\Z(T,0)}\right],
    \label{eq:excess_pressure}
\end{equation}
where $V$ is the volume of the system. From the excess pressure, the net baryon or isospin density can be calculated as
\begin{equation}
    \frac{\N(T,\mY)}{T^3} = \frac{\partial}{\partial (\mY/T)}\left[\frac{\dP(T,\mY)}{T^4}\right].
\end{equation}
Owing to the sign problem of lattice QCD, it is only possible to evaluate Eq. \eqref{eq:excess_pressure} approximately e.g. by expanding the right hand side in a Taylor series in $\mY$ and retaining terms up to some (even) order $N$ viz.
\begin{equation}
    \frac{\dP_N^T(T,\mY)}{T^4} = \sum_{n=1}^{N/2} \frac{\chi_{2n}^Y(T)}{(2n)!} \mT^{2n}.
    \label{eq:taylor_pressure}
\end{equation}
This is the $N$th order Taylor estimate of $\dP(T,\mY)$. Only even powers of $\mY$ appear in the expansion due to the particle-antiparticle symmetry of the system. The calculation of the Taylor coefficient $\chi_{2n}^Y$ requires the calculation of terms such as $\langle (D_1^Y)^a (D^Y_2)^b (D^Y_3)^c\cdots\rangle$ where
\begin{equation}
    D_n^Y(T) = \frac{\partial^n \ln \det \M(T,\mY)}{\partial (\mY/T)^n}\, \bigg\vert_{\mY=0},
    \label{eq:Dn}
\end{equation}
 $a+2b+3c+\dots = 2n$, and the angular brackets $\langle\cdot\rangle$ denote the expectation value w.r.t. an ensemble of gauge configurations generated at the same temperature $T$ but at $\mY=0$~\cite{Allton:2005gk,Gavai:2004sd}:
\begin{equation}
    \big\langle \O(T)\big\rangle = \frac{\int \mathcal{D} U \, \O(T) \, e^{-S_G(T)} \det \M(T,0)}{\int \mathcal{D} U\,e^{-S_G(T)} \det \M(T,0)}.
    \label{eq:angular_brackets}
\end{equation}

A typical lattice QCD calculation starts by calculating the first $N$ derivatives $D^Y_1,\dots,D^Y_N$ stochastically using $\Nrv \sim \O(10^2$ - $10^3)$ random volume sources per gauge configuration. With these derivatives, it is possible to calculate all the Taylor coefficients up to $\chi^Y_N$. The same derivatives however also contribute to higher-order Taylor coefficients through products such as $D^Y_ND^Y_1$, $(D^Y_N)^2$, etc. In fact, as already mentioned in Sec.~\ref{sec:introduction}, the contribution of $D^Y_1,\dots,D^Y_N$ to all orders in $\mY$ can be resummed into an exponential factor. One can thus write a resummed estimate for $\dP(T,\mY)$ as
\begin{equation}
    \frac{\dP_N^R(T,\mY)}{T^4}=\frac{\Nt^3}{\Ns^3} \ln\left[\Re\left\langle\exp \left(\sum_{n=1}^N \frac{\overline{D^Y_n}(T)}{n!}\mT^n\right)\right\rangle\right].
    \label{eq:resummed_pressure}
\end{equation}
The symbol Re in the above equation stands for the real part of a complex number. It can be proved that the $D^Y_n$ are real (imaginary) for $n$ even ($n$ odd). Hence the exponential in Eq.~\eqref{eq:resummed_pressure} is a complex quantity. For real $\mY$, the partition function is real and the imaginary part vanishes when averaged over all gauge configurations. For finite ensembles, the imaginary part can be discarded provided that it is zero within error.

The overline over $D^Y_n$ denotes the average of the $\Nrv$ stochastic estimates of $D^Y_n$. As $\Nrv\to\infty$, $\overline{D^Y_n} \to D^Y_n$ and Eq.~\eqref{eq:resummed_pressure} becomes exact. For finite $\Nrv$ however the exponential factor contains stochastic bias, which can be seen as follows: If we expand the exponential in a Taylor series, then we get terms such as $(\overline{D^Y_m})^p (\overline{D^Y_n})^q\cdots$ which contain products of estimates coming from the same random vector and are hence not truly independent estimates. Although stochastic bias can be shown to be suppressed by powers of $N_\text{rv}^{-1}$, it can still be significant depending upon the observable and the value of $\mY/T$. It therefore needs to be subtracted in order to obtain a better estimate of $\dP(T,\mY)$.

Stochastic bias is not an issue in the calculation of the Taylor coefficients, although such products also appear there, because there exist formulas for efficiently evaluating the unbiased estimate of {\it finite} products of the derivatives~\cite{Steinbrecher:2018jbv,Mitra:2022vtf}. Taking advantage of this, one way of avoiding stochastic bias is by expanding Eq.~\eqref{eq:resummed_pressure} in a cumulant expansion and retaining the first $M$ terms viz.
\begin{align}
\frac{\dP^C_{N,M}(T,\mY)}{T^4}&=\frac{\Nt^3}{\Ns^3} \sum_{m=1}^M \Re\left[\frac{\K_m\left(X^Y_N(T,\mY)\right)}{m!}\right], \notag \\
X^Y_N(T,\mY) &= \sum_{n=1}^N \frac{D^Y_n(T)}{n!}\mT^n.
\label{eq:cumulant_expansion}
\end{align}
The first four cumulants are given by
\begin{align}
    \K_1(X^Y_N) &= \langle X^Y_N \rangle, \notag \\
    \K_2(X^Y_N) &= \langle (X^Y_N)^2 \rangle - \langle X^Y_N \rangle^2, \notag \\
    \K_3(X^Y_N) &= \langle (X^Y_N)^3 \rangle - 3 \langle X^Y_N \rangle \langle (X^Y_N)^2 \rangle + 2 \langle X^Y_N \rangle^3, \notag \\
    \K_4(X^Y_N) &= \langle (X^Y_N)^4 \rangle - 4 \langle X^Y_N \rangle \langle (X^Y_N)^3 \rangle - 3 \langle (X^Y_N)^2 \rangle^2 \notag \\
            &\phantom{=}+ 12 \langle (X^Y_N)^2 \rangle \langle X^Y_N \rangle^2 - 6 \langle X^Y_N \rangle^4.
\label{eq:first_four_cumulants}
\end{align}
However, as we have already noted, with this approach both all-orders resummation as well as knowledge of the phase factor are lost. Therefore in this paper, instead of expanding the resummed pressure we propose to modify the argument of the exponential factor so that the stochastic bias is subtracted up to a certain order of either the Taylor or the cumulant expansion. Although the bias is subtracted on a configuration-by-configuration basis, the resulting expression for $\dP(T,\mY)$ too can be shown to be free of stochastic bias up to the same order (Appendix~\ref{app:proof}).

We begin with the Taylor series case first. The analog of Eq.~\eqref{eq:resummed_pressure}, but with the exponential unbiased to $\O(\mY^N)$, is achieved by replacing $\overline{D^Y_n}(T)$ by $\C^Y_n(T)$ i.e.
\begin{widetext}
\begin{equation}
    \frac{\dP^{R(\text{unb})}_N(T,\mY)}{T^4} = \frac{\Nt^3}{\Ns^3} \, \ln\Bigg[\Re \left\langle \exp \left(\sum_{n=1}^N \frac{\C^Y_n(T)}{n!}\mT^n\right)\right\rangle\Bigg],
\label{eq:unbiased_resummed-1}
\end{equation}
where the $\C^Y_n(T)$ for $1 \leq n \leq 4$ are given by 
\begin{align}
    &\C^Y_1 = \overline{D^Y_1}, \notag \\
    &\C^Y_2 = \overline{D^Y_2} + \left(\overline{(D^Y_1)^2} - \left(\overline{D^Y_1}\right)^2\right), \notag \\
    &\C^Y_3 = \overline{D^Y_3} + 3\left(\overline{D^Y_2 D^Y_1} - \overline{D^Y_2}\;\overline{D^Y_1}\right) + 
           \left(\overline{(D^Y_1)^3} - 3\,\overline{(D^Y_1)^2}\;\overline{D^Y_1} + 2\,\left(\overline{D^Y_1}\right)^3\right), \notag \\
    &\C^Y_4 = \overline{D^Y_4} + 3\left(\overline{(D^Y_2)^2} - \left(\overline{D^Y_2}\right)^2\right)+ 4\left(\overline{D^Y_3 D^Y_1} - \overline{D^Y_3}\;\overline{D^Y_1}\right) + 6\left( \overline{D^Y_2 (D^Y_1)^2} - \overline{D^Y_2}\;\overline{(D^Y_1)^2}\right) - 3\,(\overline{(D^Y_1)^2})^2 \notag \\
    &\phantom{\C^Y_4} - 12 \left(\overline{D^Y_2 D^Y_1}\;\overline{D^Y_1} - \overline{D^Y_2}\left(\overline{D^Y_1}\right)^2\right) +
     \overline{(D^Y_1)^4} - 4\,\,\overline{(D^Y_1)^3}\;\overline{D^Y_1} + 12\,\overline{(D^Y_1)^2}\left(\overline{D^Y_1}\right)^2 - 6\left(\overline{D^Y_1}\right)^4,\quad\text{etc.}
\label{eq:Cn}
\end{align}
\end{widetext}
 The first term in each equation is just $\overline{D^Y_n}$. The remaining terms are the ``counterterms'' that are added to subtract the stochastic bias. A term such as $\overline{D^Y_2D^Y_1}$ in the above equations stands for the unbiased product of $D^Y_2$ and $D^Y_1$. Similarly, $\overline{(D^Y_1)^2}$ represents the unbiased square of $D^Y_1$. By contrast, a term such as $(\overline{D^Y_1})^2$ represents the biased square i.e. the square of the average of $D^Y_1$. 
The exponential constructed in this way is unbiased to $\O(\mY^N)$. We will prove in Appendix~\ref{app:proof} that both the Taylor expansion of the exponential as well as the excess pressure calculated from it (Eq.~\eqref{eq:unbiased_resummed-1}) are free of stochastic bias up to the same order.

As already noted, the first term in each $\C^Y_n$ is simply $\overline{D^Y_n}$. In the limit $\Nrv \to \infty$, this term approaches the correct value of $D^Y_n$. The rest of the terms for each $\C^Y_n$ also cancel each other out as $\Nrv\to\infty$, since in that limit the distinction between biased and unbiased products vanishes. Thus $\C^Y_n\to D^Y_n$ as $\Nrv\to\infty$ and hence Eq.~\eqref{eq:unbiased_resummed-1} too represents an all-orders resummation of the derivatives $D^Y_1,\dots,D^Y_N$, the only difference this time being that the stochastic bias is eliminated to $\O(\mY^N)$.

Although Eq.~\eqref{eq:unbiased_resummed-1} is an improvement over Eq.~\eqref{eq:resummed_pressure}, it is possible to do still better. In a typical lattice QCD calculation, each stochastic estimate of $D^Y_1,\dots,D^Y_N$ is constructed using the same random source. Therefore, the different stochastic estimates can be actually thought of as different estimates of the operator $X^Y_N(T,\mY)$, where $X^Y_N(T,\mY)$ is as given in Eq.~\eqref{eq:cumulant_expansion}. It is possible to write a version of Eq.~\eqref{eq:resummed_pressure} in which the bias is eliminated up to a certain power of $X^Y_N$ itself, by writing
\begin{widetext}
\begin{equation}
    \frac{\dP^{R(\text{unb})}_{N,M}(T,\mY)}{T^4} = \frac{\Nt^3}{\Ns^3} \, \ln\left[\Re\left\langle \exp \left(\sum_{m=1}^M \frac{\L_m(X^Y_N(T,\mY))}{m!}\right)\right\rangle\right],
\label{eq:unbiased_resummed-2}
\end{equation}
where
\begin{align}
   \L_1 &= \overline{X^Y_N}, \notag \\
   \L_2 &= \overline{(X^Y_N)^2} - \big(\overline{X^Y_N}\big)^2, \notag \\
   \L_3 &= \overline{(X^Y_N)^3} - 3\,\big(\overline{X^Y_N}\big)\;\big(\overline{(X^Y_N)^2}\big) + 2\,\big(\overline{X^Y_N}\big)^3, \notag \\
     \L_4 &= \overline{(X^Y_N)^4}- 4\,\big(\overline{(X^Y_N)^3}\big)\;\big(\overline{X^Y_N}\big) - 3\,\big(\overline{(X^Y_N)^2}\big)^2 
         + 12\,\big(\overline{X^Y_N}\big)^2\;\big(\overline{(X^Y_N)^2}\big) - 6\,\big(\overline{X^Y_N}\big)^4,\quad\text{etc.}
\label{eq:Lm}
\end{align}
\end{widetext}
We note that Eqs.~\eqref{eq:Lm} resemble the cumulant formulas Eqs.~\eqref{eq:first_four_cumulants}, but with two differences:
\begin{enumerate}
\item[(i)] The expansion is in the space of all random estimates for a single gauge configuration rather than in the space of all gauge configurations.
\item[(ii)] The powers $(X^Y_N)^p$ are replaced by their respective unbiased estimates $\overline{(X^Y_N)^p}$. 
\end{enumerate}
In the limit $\Nrv\to\infty$, the difference between biased and unbiased estimates vanishes. Then the $\L_m$ are just the cumulants of $X^Y_N$ over the set of all random estimates for a single gauge configuration. In the double limit $M\to\infty$ and $\Nrv\to\infty$ therefore, the argument of the exponential in Eq.~\eqref{eq:unbiased_resummed-2} is just the cumulant expansion of $\overline{\;e^{X^Y_N}\;}$. This observation helps to clarify the meaning of bias subtraction: It is the systematic (order-by-order) replacement of the incorrect (biased) estimate $e^{\,\overline{X^Y_N}}$ of the exponential factor by the correct estimate $\overline{\;e^{X^Y_N}\;}$.

In addition to the excess pressure and the number density, we have also presented results for the average phase factor. As already mentioned, the $D^Y_n$ are real (imaginary) for even $n$ (for odd $n$) and hence the exponential factor is complex even when $\mB$ is real~\footnote{For finite isospin, the odd $D^Y_n$ are identically zero and hence the exponential is real for both real and imaginary $\mI$. For complex $\mI$ however, the phase factor will also be complex for the isospin case.}. Although its imaginary part vanishes, the real part still receives a contribution $\cos \Theta(T,\mB)$ at $\mB\ne0$ from the phase of the exponential. The average phase factor $\langle\cos\Theta(T,\mB)\rangle$ is a measure of the difficulty of the calculation at finite $\mB~$\footnote{This is true not just for the baryochemical potential $\mB$ but for any chemical potential for which there is a sign problem e.g. $\mu_S$.}. As $\mB$ is increased, $\langle\cos\Theta(T,\mB)\rangle\to0$ and the rapid fluctuations of the phase factor cause the calculation to break down. This happens as $\mB\to\vert\mu_B^c\vert$, where $\mu^c_B$ is the nearest singularity to $\mB=0$ of the QCD partition function in the complex $\mB$ plane. Unlike a finite Taylor series therefore, the resummation calculation cannot be carried out to arbitrarily large $\mB$.

Similar to the $D^Y_n$, it can be shown that the $\C^Y_n$ (Eq.~\eqref{eq:unbiased_resummed-1}) too are real (imaginary) for even (odd) $n$. Similarly, the $\L_m$ (Eq.~\eqref{eq:unbiased_resummed-2}) too are real (imaginary) for even (odd) $m$ when $\mY$ is real. Hence in each case we can define an average phase factor $\langle\cos\Theta(T,\mY)\rangle$, where $\Theta(T,\mY)$ is defined as
\begin{subequations}
\begin{align}
\Theta^R_N(T,\mY) &= \Im\left[\sum_{n=1}^N \frac{D^Y_n(T)}{n!}\mT^n\right],\label{eq:phase_factor_biased}\\
\Theta^{R\text{(unb)}}_N(T,\mY) &= \Im\left[\sum_{n=1}^N \frac{\C^Y_n(T)}{n!}\mT^n\right],\label{eq:phase_factor_mu_basis}\\
\Theta^{R\text{(unb)}}_{N,M}(T,\mY) &=\Im\left[\sum_{n=1}^M \frac{\L_n(X^Y_N(T,\mY))}{n!}\right],\label{eq:phase_factor_cumulant_basis}
\end{align}
\label{eq:phase_factor}
\end{subequations}
where Im stands for the imaginary part of the argument. For real $\mY$, this is simply the sum over odd $n$. However, when written as above, the formulas are also valid for the more general case of complex $\mY$. Note that it is not possible to define a phase factor for the Taylor series. An approximation to the phase factor may be constructed by Taylor-expanding Eqs.~\eqref{eq:phase_factor} to a particular order. However the approximation diverges as $\mY$ is increased and hence it cannot be used to determine the breakdown of the calculation.

\section{Results}
\label{sec:results}
To verify our formalism, we made use of the data generated by the HotQCD collaboration~\footnote{A complete description of the gauge ensembles and scale setting can be found in Ref.~\cite{Bollweg:2021vqf}.} for its ongoing Taylor expansion calculations of the finite density QEOS, chiral crossover temperature and conserved charge cumulants at finite density~\cite{Bazavov:2017dus,HotQCD:2018pds,Bollweg:2021vqf,Bollweg:2022rps,Bollweg:2022fqq}. For these calculations, $\O(10^4$ - $10^6)$~$2$+$1$-flavor gauge configurations were generated in the temperature range $135$~MeV~$\lesssim~T~\lesssim$~$176$~MeV using a Symanzik-improved gauge action and the Highly Improved Staggered Quark (HISQ) fermion action with $\Nt=8$, $12$ and $16$ and $\Ns=4\Nt$~\cite{Follana:2006rc,Bazavov:2011nk}. The temperature for each $\Nt$ was varied by varying the lattice spacing $a$ through the gauge coupling $\beta$, and for each lattice spaing the bare light and strange quark masses $m_l(a)$ and $m_s(a)$ were also tuned so that the pseudo-Goldstone pion and kaon masses were equal to the physical pion and kaon masses respectively. The scale was determined using both the Sommer parameter $r_1$ and the kaon decay constant $f_K$. The temperature values quoted in this paper are from the $f_K$ scale.

To calculate the Taylor coefficients, on each gauge configuration the first eight derivatives $D_1^f,\dots,D_8^f$ for each quark flavor $f$ were estimated stochastically using $2000$ Gaussian random volume sources for $D^f_1$ and $500$ sources for the higher derivatives for both $\mB$ and $\mI$. The exponential-$\mu$ formalism~\cite{Hasenfratz:1983ba} was used to calculate the first four derivatives while the linear-$\mu$ formalism~\cite{Gavai:2011uk,Gavai:2014lia} was used to calculate the higher derivatives. Using this data, we calculated the excess pressure and number density for both real and imaginary baryon as well as isospin chemical potentials $\mu_B$ and $\mI$, in the range $0 \leqslant \lvert \mu_{B,I}/T \rvert \leqslant 2$, using 100k (20k) configurations per temperature for the baryon (isospin) case. Our results were obtained on $\Nt=8$ lattices for three temperatures viz. $T \sim 157$, $176$ and $135$ MeV. These temperatures were chosen as being approximately equal to $\Tpc$ and $\Tpc\pm20$~MeV, where $\Tpc=156.5(1.5)$~MeV is the chiral crossover temperature at $\mu_B=0$~\cite{HotQCD:2018pds}.

\subsection{Results for Finite Isospin Chemical Potential}
\label{ssec:isospin_chempot}
\begin{figure*}
\begin{minipage}[t]{0.99\textwidth}
\includegraphics[width=0.40\textwidth]{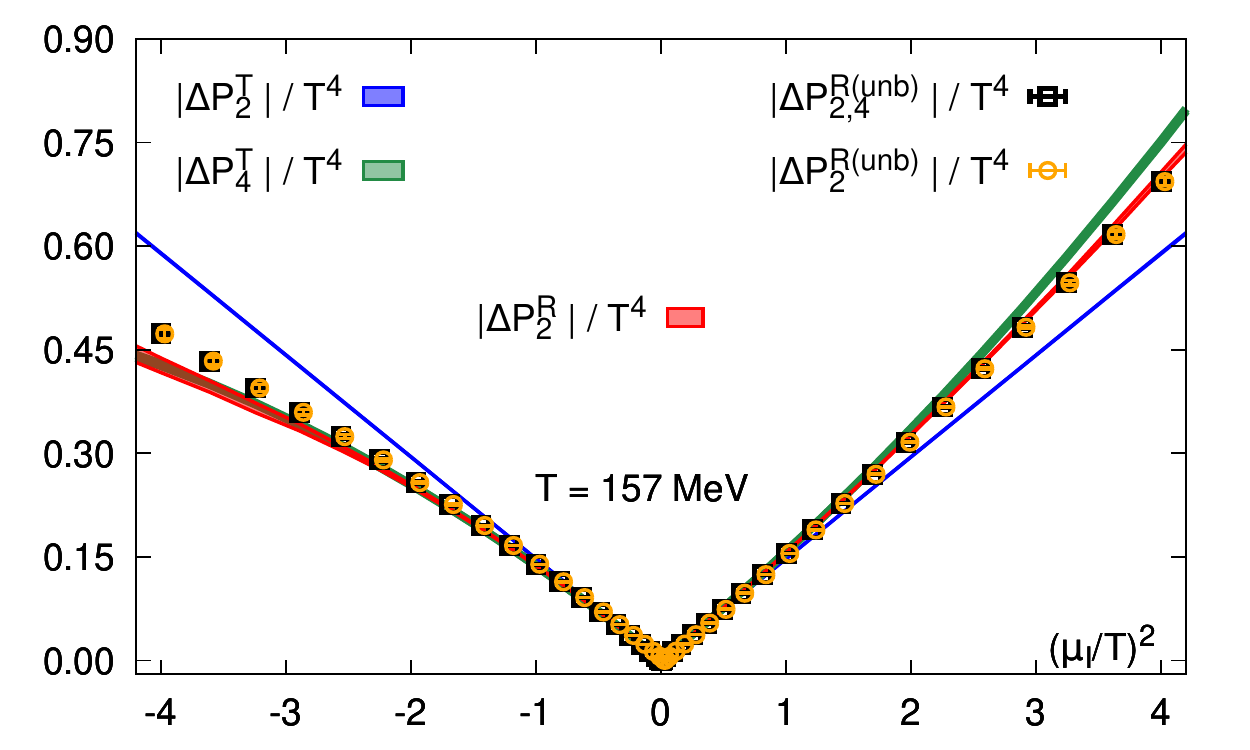}
\hspace{0.05\textwidth}%
\includegraphics[width=0.40\textwidth]{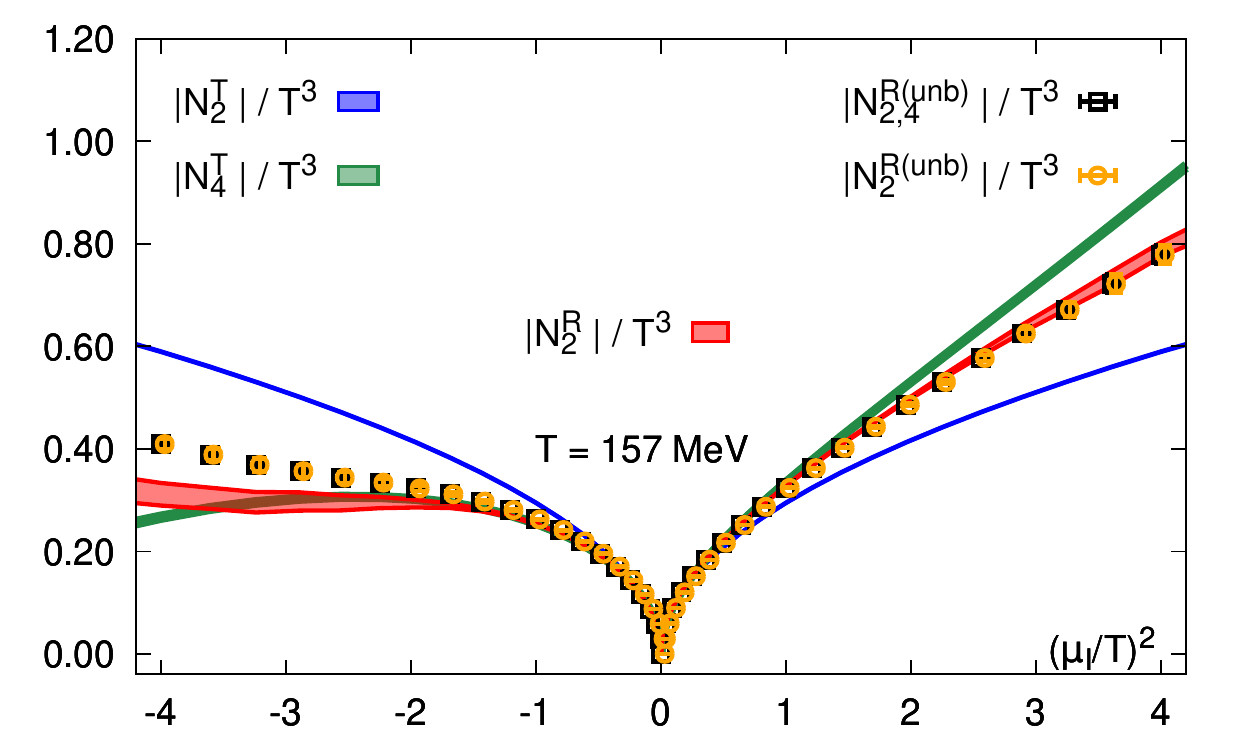}
\vspace{0.01\textheight}%
\includegraphics[width=0.40\textwidth]{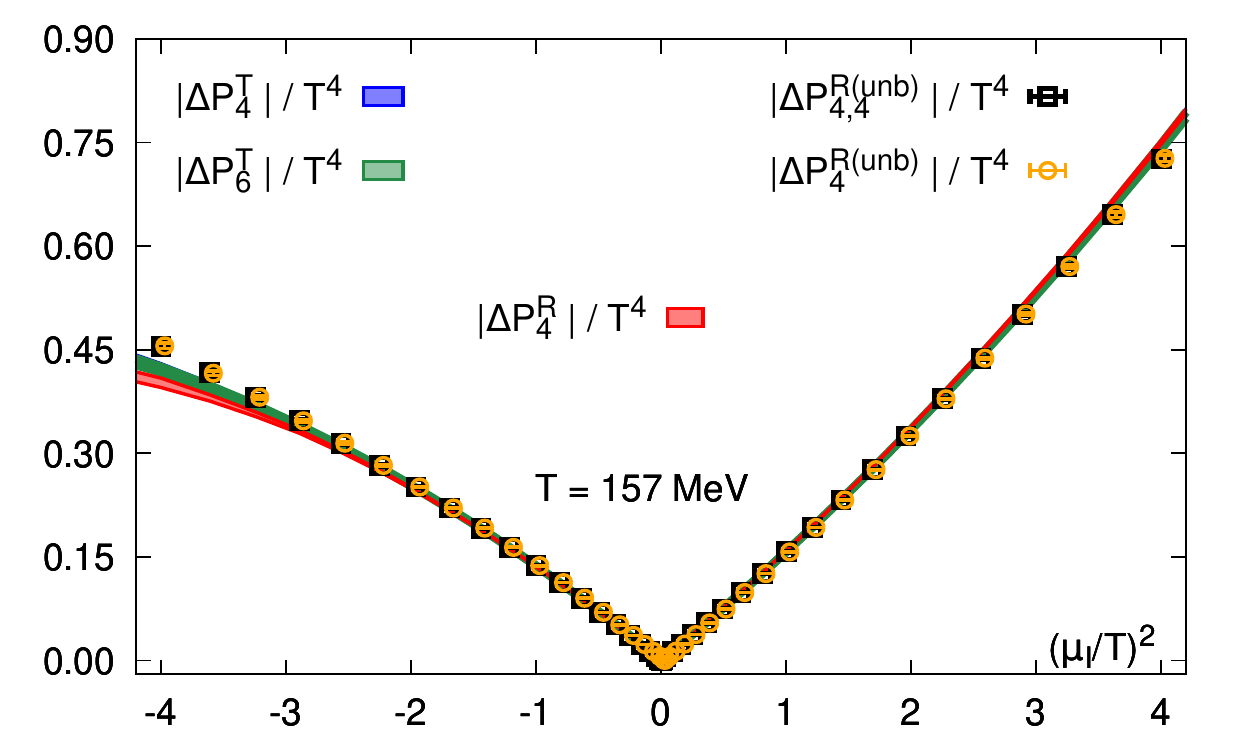}%
\hspace{0.05\textwidth}
\includegraphics[width=0.40\textwidth]{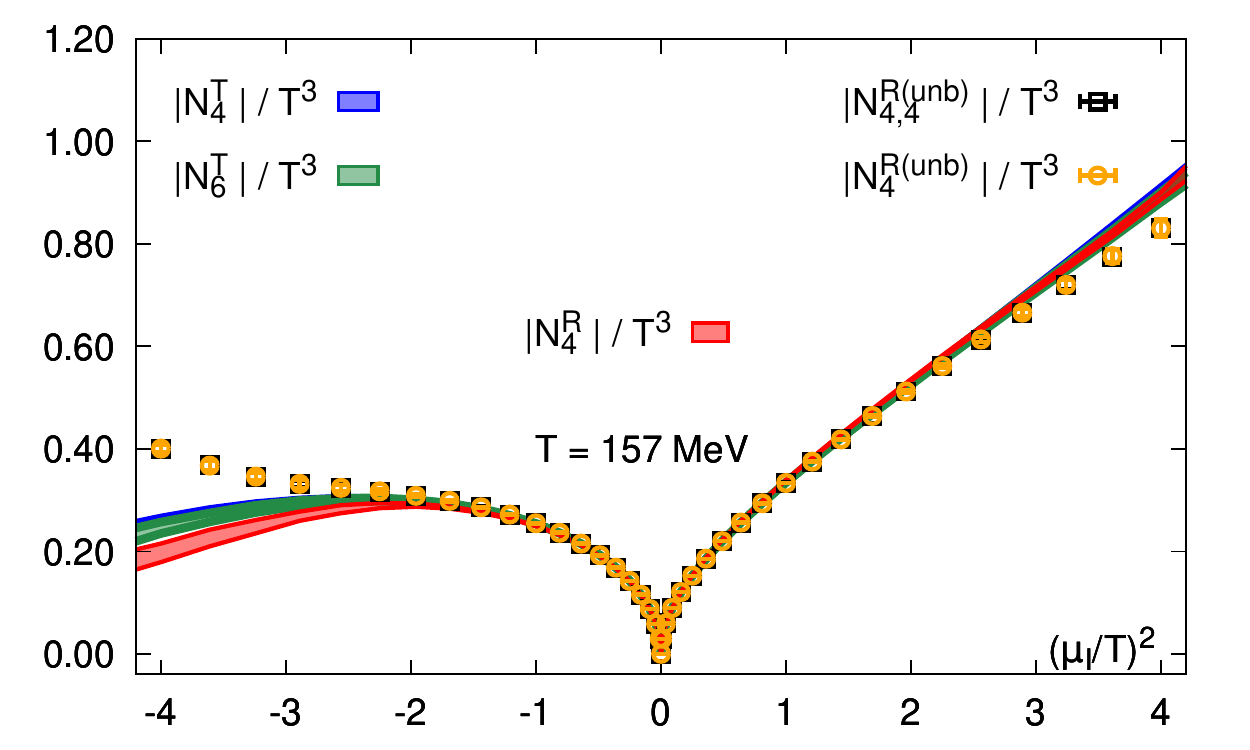}
\caption{$\dP(T,\mI)/T^4$ and $\N(T,\mI)/T^3$, calculated for $T=157$~MeV using second and fourth order biased (red bands) and unbiased resummations. Unbiased resummation results in cumulant (chemical potential) bases are plotted as black squares (orange circles); different ordered Taylor expansion results are plotted in green and blue bands respectively.\label{fig:muI-157MeV}}
\end{minipage}
\end{figure*}

Before considering the finite $\mu_B$ case, we shall first present our results for the simpler case of finite isospin chemical potential $\mI$~\cite{Son:2000xc,Brandt:2017oyy,Adhikari:2020kdn}. For finite $\mI$, the fermion determinant is real and hence there is no sign problem. Hence direct simulations of the system are possible unlike in the $\mB$ case. As a result, both Taylor expansion of observables as well as resummation of the Taylor series are unnecessary. Our reason for studying the isospin case is that the absence of the sign problem makes it possible to calculate observables up to much larger values of $\hmI$ compared to the $\mu_B$ case, and it is precisely for these values that bias can become significant. The isospin calculations thus allow for a more stringent test of the formalism.

We present our results for $\dP/T^4$ and $\N/T^3$ for $T=157$~MeV, resummed to second (fourth) order using the different resummation formulas: Eq.~\eqref{eq:resummed_pressure} (red bands), Eq.~\eqref{eq:unbiased_resummed-1} (orange circles) and Eq.~\eqref{eq:unbiased_resummed-2} (black squares), in the top (bottom) two plots of Fig.~\ref{fig:muI-157MeV}. In each of the plots, we also plot the Taylor expansion results (blue and green bands) for purposes of comparison.

We find that the fourth order Taylor results differ from the second order results for $|\hmI^2|\gtrsim1$. Turning next to the resummation results, we find that the biased resummation results agree well overall with the fourth order Taylor results for both real as well as imaginary chemical potentials. The resummation results were obtained by resumming the derivative $D^I_2$ while the fourth order Taylor results also contain contributions from $D^I_4$~\footnote{Note that $D^I_1$ and $D^I_3$ are identically zero.}. The agreement between these two results would therefore suggest that the latter two derivatives do not contribute significantly for $0\leqslant|\hmI^2|\leqslant4$. Before arriving at this conclusion however, it is necessary to account for the stochastic bias that is present in the results of Eq.~\eqref{eq:resummed_pressure}. In fact, the unbiased resummation results, obtained using either Eq.~\eqref{eq:unbiased_resummed-1} or Eq.~\eqref{eq:unbiased_resummed-2}, lie in between the second and fourth order Taylor results. Moreover the results from Eq.~\eqref{eq:unbiased_resummed-1} and Eq.~\eqref{eq:unbiased_resummed-2} are practically identical, which means that it is sufficient to eliminate bias to $\O(\mu_I^2)$ for the range of chemical potentials considered here. We conclude that the derivatives $D^I_3$ and $D^I_4$ do in fact contribute at fourth order, and that the biased resummation results will approach the unbiased results in the limit $\Nrv\to\infty$.

Subtracting bias becomes important at higher orders because the lower order derivatives contribute through higher powers e.g. the derivative $D^I_2$ contributes at sixth order via $(D^I_2)^3$ respectively. In the lower two plots of Fig.~\ref{fig:muI-157MeV}, we compare results from fourth order resummations with fourth and sixth order Taylor expansion results. The sixth order results differ only slightly from the fourth order results for both $\dP/T^4$ as well as $\N/T^3$ over the entire range $-4\leqslant\hmI^2\leqslant4$. By contrast, the biased resummation results differ significantly from both fourth and sixth order Taylor results and are in fact non-monotonic for $\N/T^3$ for imaginary $\mu_I$. Subtracting the bias to $\O(\mI^4)$ yields results that are in very good agreement with the sixth order Taylor result. No further changes result if the bias is further subtracted up to fourth order of the cumulant expansion.

\subsection{Results for Finite Baryon Chemical Potential}
\label{ssec:baryon_chempot}
\begin{figure*}
\begin{minipage}[!t]{0.99\textwidth}
\includegraphics[width=0.40\textwidth]{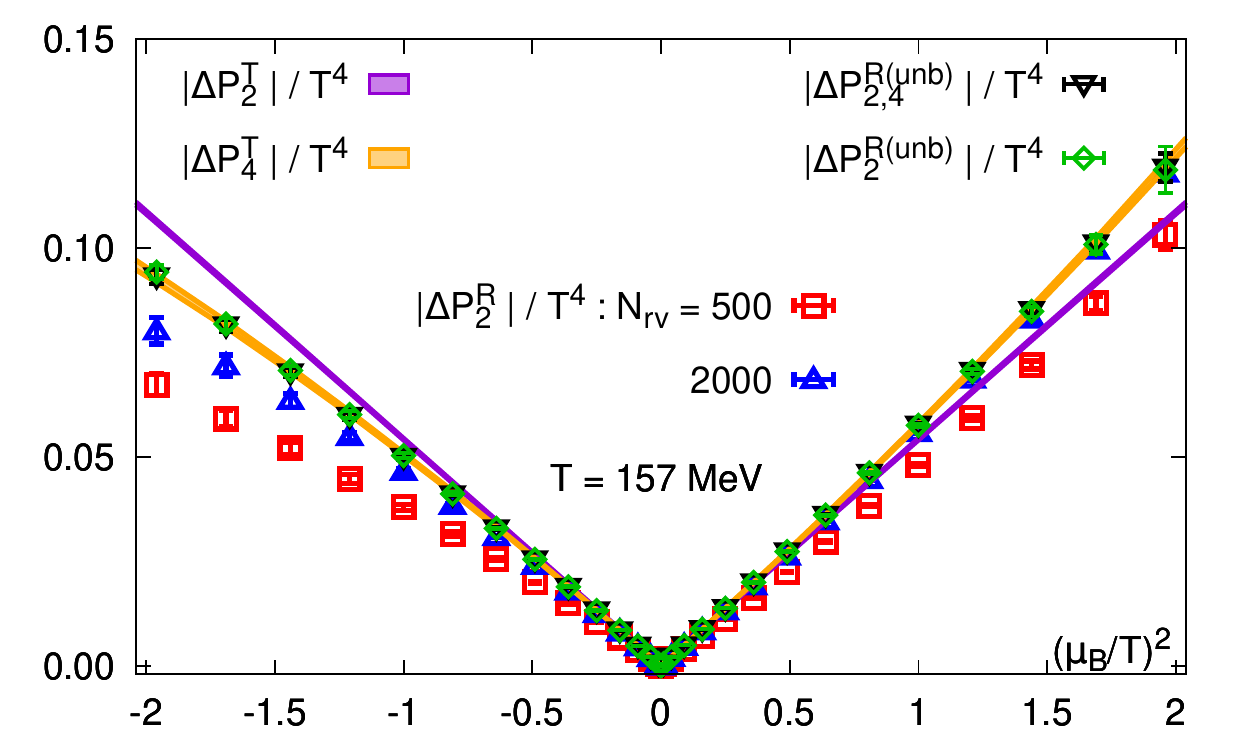}%
\hspace{0.05\textwidth}%
\includegraphics[width=0.40\textwidth]{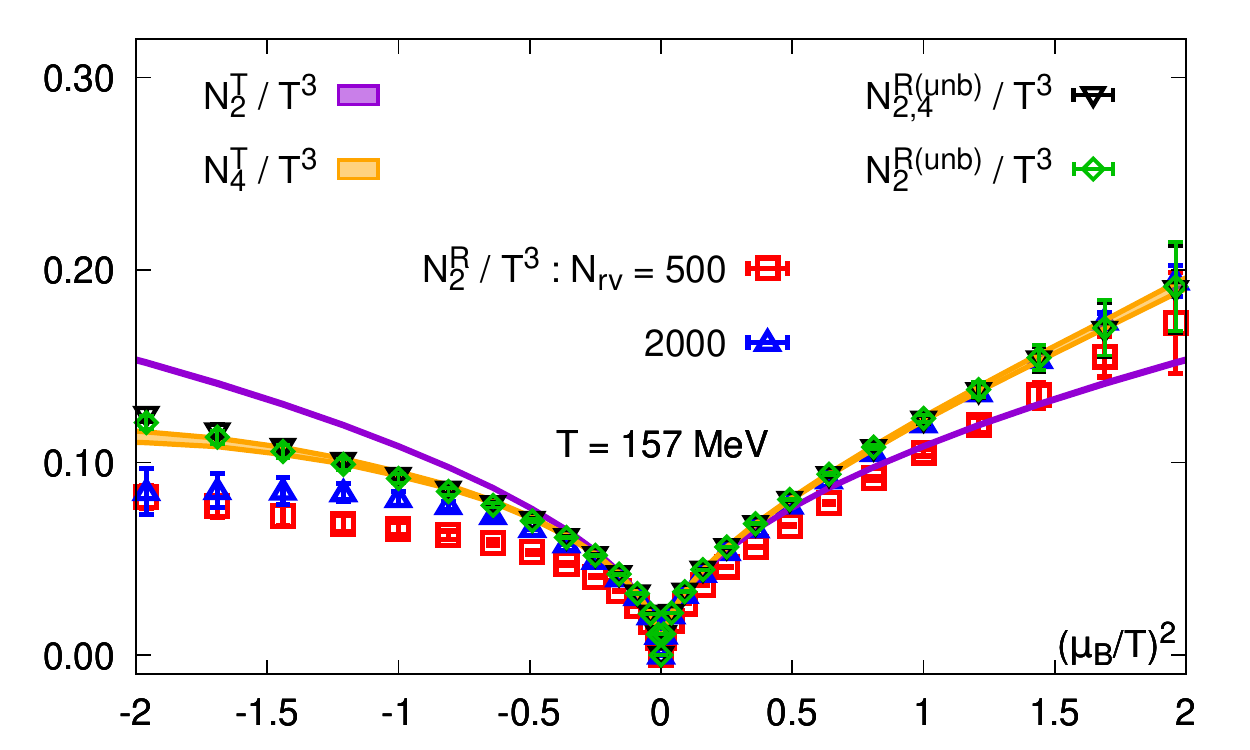}
\vspace{0.01\textheight}%
\includegraphics[width=0.40\textwidth]{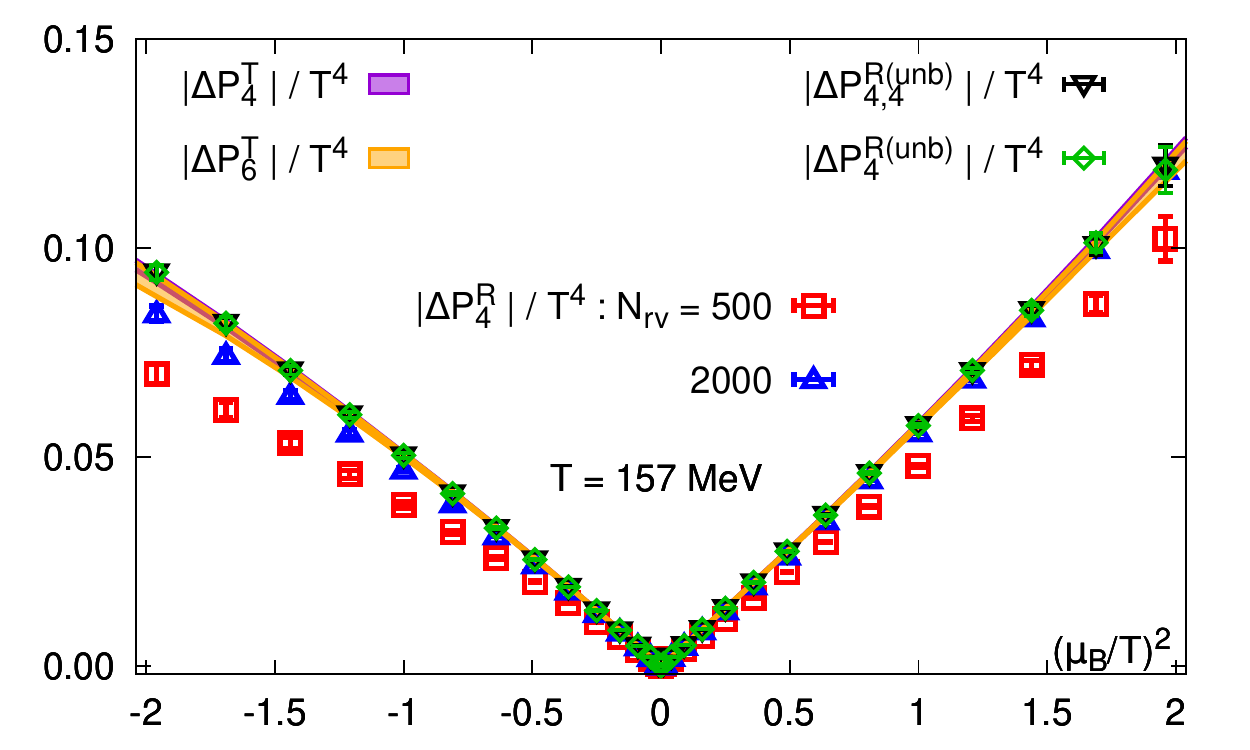}%
\hspace{0.05\textwidth}%
\includegraphics[width=0.40\textwidth]{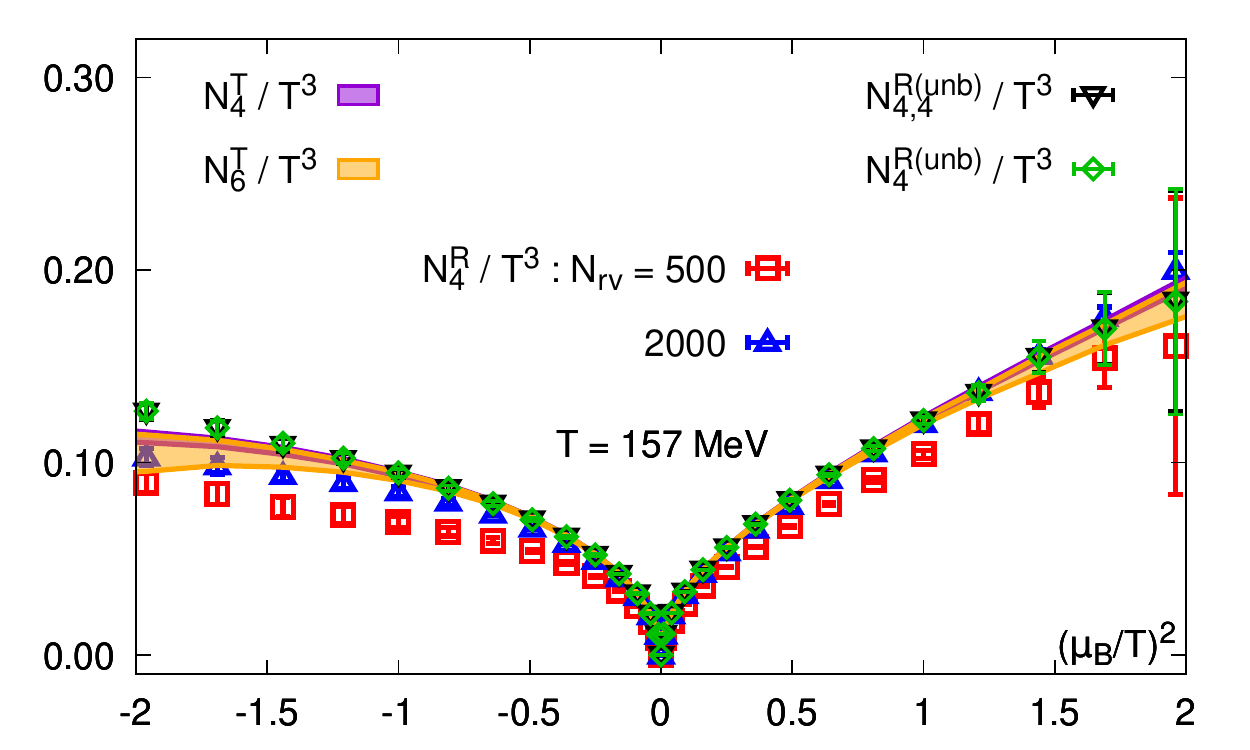}
\caption{$\dP(T,\mB)/T^4$ and $\N(T,\mB)/T^3$, calculated for $T=157$~MeV using second and fourth order biased and unbiased resummations and second, fourth and sixth order Taylor expansions. The Taylor expansion results are plotted as purple and orange bands, whereas unbiased resummation results for cumulant (chemical potential) bases are presented as black inverted triangles (green diamonds). The biased results for $500$ and $2000$ random sources are shown as red squares and blue triangles respectively.\label{fig:muB-157MeV}}
\end{minipage}
\end{figure*}

The resummed results for the QEOS at finite baryochemical potential $\mB$ have been previously presented in Ref.~\cite{Mondal:2021jxk}. Those results were obtained using the biased formula Eq.~\eqref{eq:resummed_pressure}, but using the full set of $2000$ independent random estimates for $D^B_1$. The use of $2000$ stochastic estimates instead of the usual $500$ does decrease the stochastic bias, however it does not subtract the contribution to the bias coming from the higher order derivatives. By contrast, the unbiased exponential formulas treat all $N$ derivatives on an equal footing and subtract all  contributions to the bias up to a certain order. The results we will present here will show that the unbiased exponential is able to achieve a greater reduction of the stochastic bias despite working with only $\Nrv=500$ stochastic estimates of the derivatives $D^B_1,\dots,D^B_N$.

We present our results for $\dP(T,\mB)$ and $\N(T,\mB)$ for $T=157$~MeV in Fig.~\ref{fig:muB-157MeV}. The resummation results were calculated using both the biased (Eq.~\eqref{eq:resummed_pressure}) as well as the unbiased exponential (Eqs.~\eqref{eq:unbiased_resummed-1} and \eqref{eq:unbiased_resummed-2}) (green diamonds and black inverted triangles respectively). Furthermore, the biased resummation results were calculated using both $\Nrv=500$ (red squares) and $\Nrv=2000$ (blue triangles) for the operator $D_1^B$. In all four plots, we have also compared the resummation results to Taylor expansion results (purple and orange bands) as well.

In the upper two plots of Fig.~\ref{fig:muB-157MeV}, we compare the second order resummation results with second and fourth order Taylor expansion results. We find that although the biased resummation results calculated using $\Nrv=500$ random sources agree with the second order Taylor results for $\dP(T,\mB)$  for real $\mB$, in all other cases they differ from the second and even from the fourth order Taylor results. When the same biased results are recalculated using $\Nrv=2000$ random estimates for $D^B_1$ this difference decreases, proving that the discrepancy is in fact due to stochastic bias. In fact, even for $\dP^R_2(T,\mB)$ for real $\mB$, the results recalculated this way move away from the second order results and instead agree with the fourth order Taylor results. By contrast the unbiased resummation results always agree with the fourth order Taylor expansion results, even though the resummation was only carried out for the derivative $D^B_2$. Also, the agreement between the results of Eq.~\eqref{eq:unbiased_resummed-1} and Eq.~\eqref{eq:unbiased_resummed-2} prove that it is sufficient to eliminate bias to $\O(\hmB^2)$ for the two observables and for the range of chemical potentials considered here. It is also clear from the figures that the biased results will approach the unbiased results as $\Nrv$ is increased. Note however that the latter were calculated using only $\Nrv=500$ stochastic estimates. Hence the unbiased results clearly converge faster to the $\Nrv\to\infty$ limit as compared to the biased results. The fourth order resummation results too present a similar picture, as can be seen from the lower two plots of Fig.~\ref{fig:muB-157MeV}.

\begin{figure*}
\begin{minipage}[!t]{0.99\textwidth}
\includegraphics[width=0.40\textwidth]{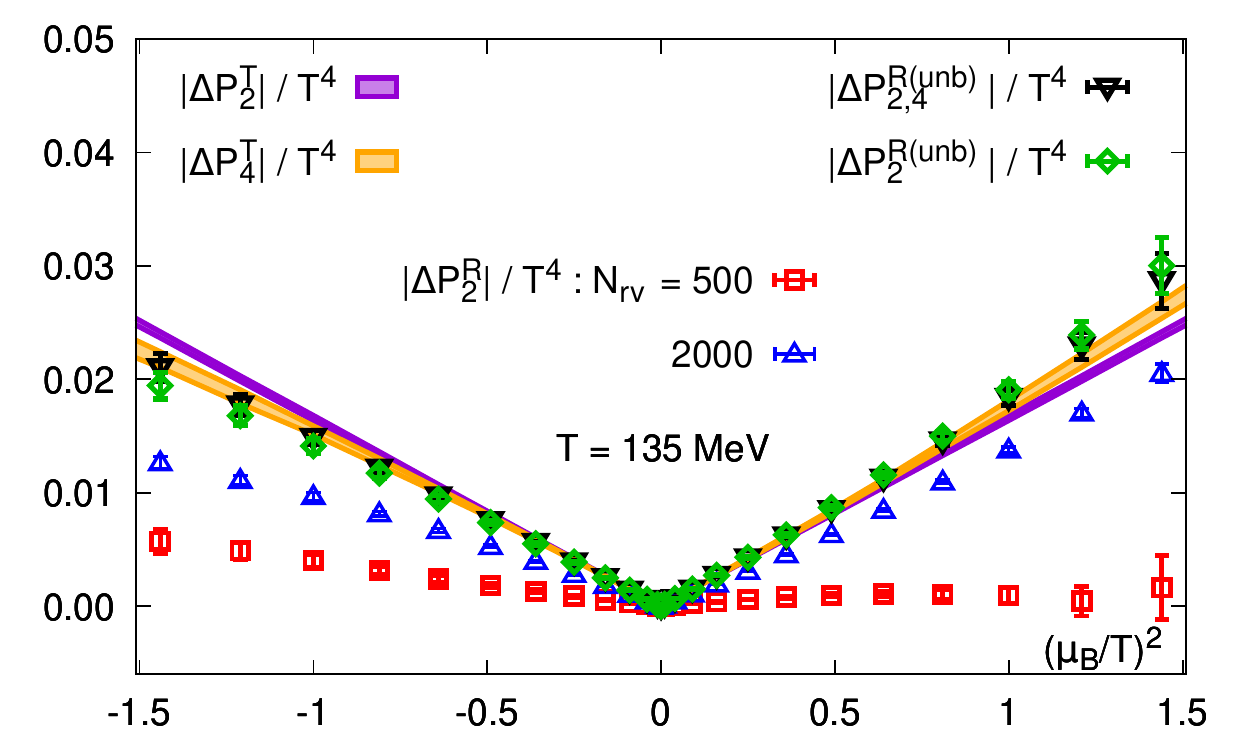}%
\hspace{0.05\textwidth}%
\includegraphics[width=0.40\textwidth]{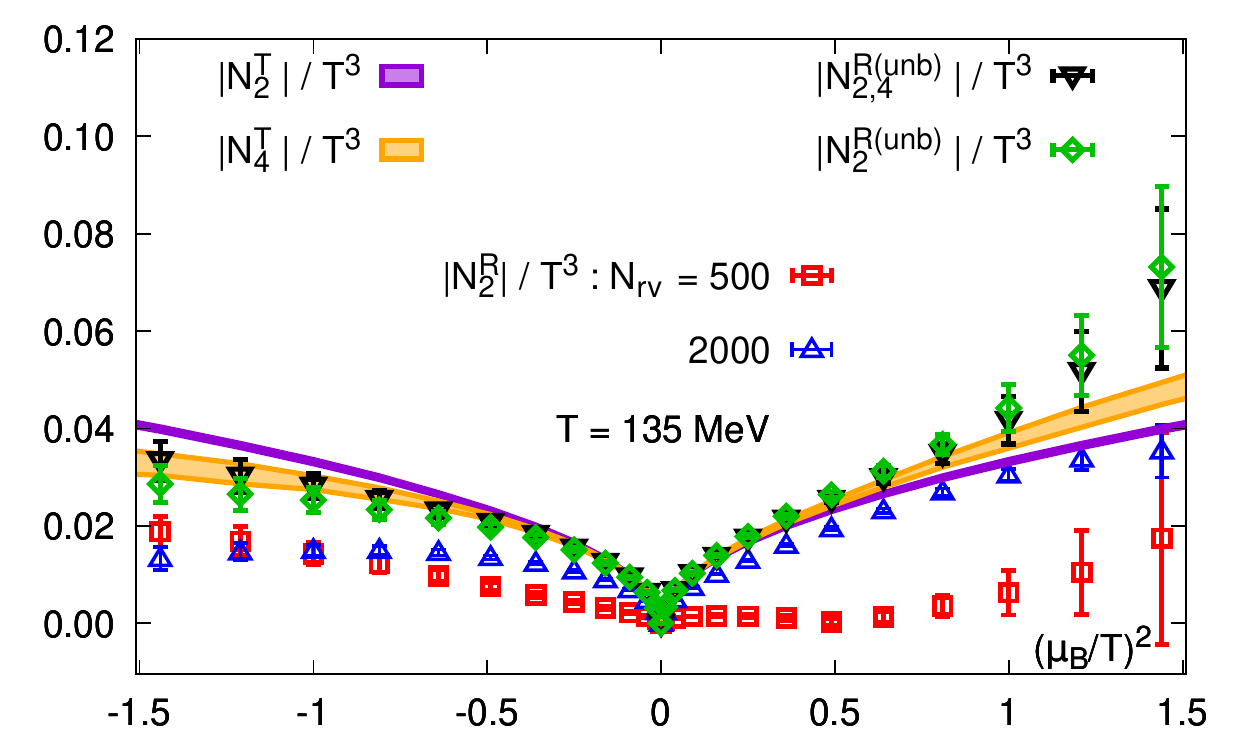}
\vspace{0.01\textheight}%
\includegraphics[width=0.40\textwidth]{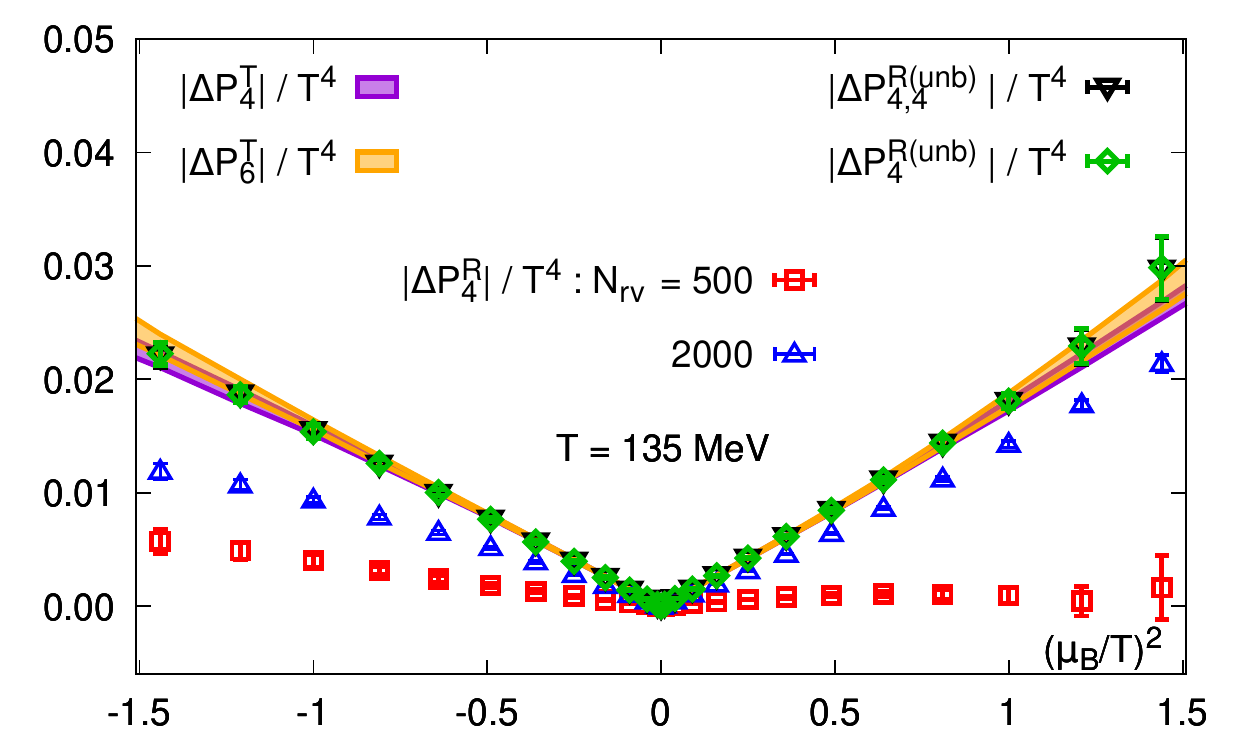}%
\hspace{0.05\textwidth}%
\includegraphics[width=0.40\textwidth]{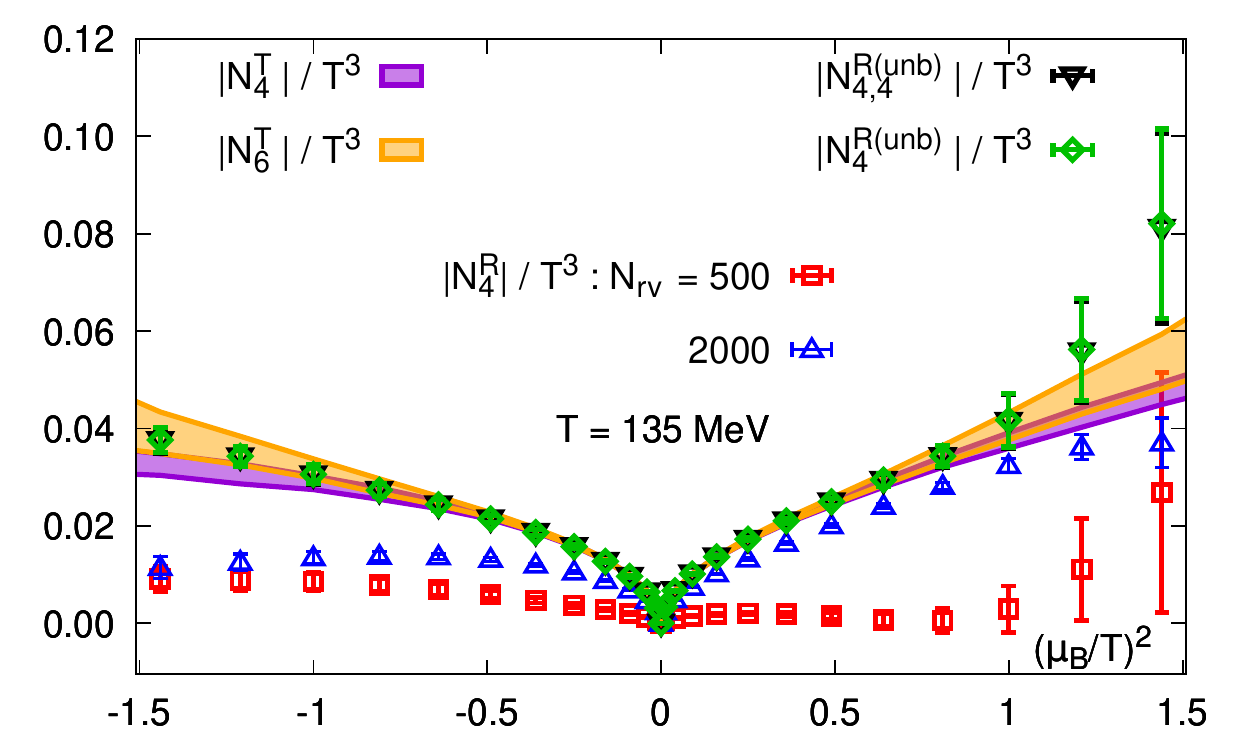}
\caption{$\dP(T,\mB)/T^4$ and $\N(T,\mB)/T^3$ for $T=135$~MeV using Taylor expansion and biased and unbiased resummation. All colors and symbols are the same as in Fig.~\ref{fig:muB-157MeV}.}
\label{fig:muB-135MeV}
\end{minipage}
\end{figure*}

The difference between biased and unbiased resummation becomes significant as one goes to lower temperatures. In Fig.~\ref{fig:muB-135MeV}, we present the resummation results for $\dP(T,\mB)$ and $\N_B(T,\mB)$ for $T=135$~MeV. The red squares are the biased results obtained using Eq.~\eqref{eq:resummed_pressure} with 500 stochastic estimates for $D^B_1$. The blue triangles were obtained using the same approach but with 2000 stochastic estimates for $D^B_1$. We see that the former results are close to zero while the latter are clearly non-zero and closer to the Taylor expansion results. The results thus indicate the presence of stochastic bias which needs to be subtracted before genuine higher order contributions can be identified.

We note that the fourth order Taylor expansion results only slightly correct the second order results over the entire range of $\hmB$. The higher order contributions of the operator $D_2^B$ are thus expected to be small. Indeed, the unbiased resummation results, whether obtained using Eq.~\eqref{eq:unbiased_resummed-1} or Eq.~\eqref{eq:unbiased_resummed-2}, are in very good agreement with the fourth order Taylor series for all chemical potentials. Moreover, the good agreement between the two approaches suggests that it suffices to subtract the bias to $\O(\hmB^2)$ for $\vert\hmB^2\vert\leqslant 1.5$.

We also note that the unbiased results were obtained using only 500 stochastic estimates for $D^B_1$ and $D^B_2$. While Eqs.~\eqref{eq:unbiased_resummed-1} or \eqref{eq:unbiased_resummed-2} are more complicated to evaluate than Eq.~\eqref{eq:resummed_pressure}, this calculational cost must be compared to the cost of calculating and storing several extra random volume source estimates of $D^B_1$ for each of $\O(10^5$~-~$10^6)$ gauge configurations. Similarly, while it is also possible to avoid stochastic bias by computing the $D^B_n$ exactly~\cite{Borsanyi:2022soo}, the method is expensive and does not scale easily to the lattice volumes considered in this study. For these reasons, we believe that it is advantageous to always use the unbiased exponential for exponential resummation of the Taylor series.

\begin{figure*}
\begin{minipage}[t]{0.99\textwidth}
\includegraphics[width=0.40\textwidth]{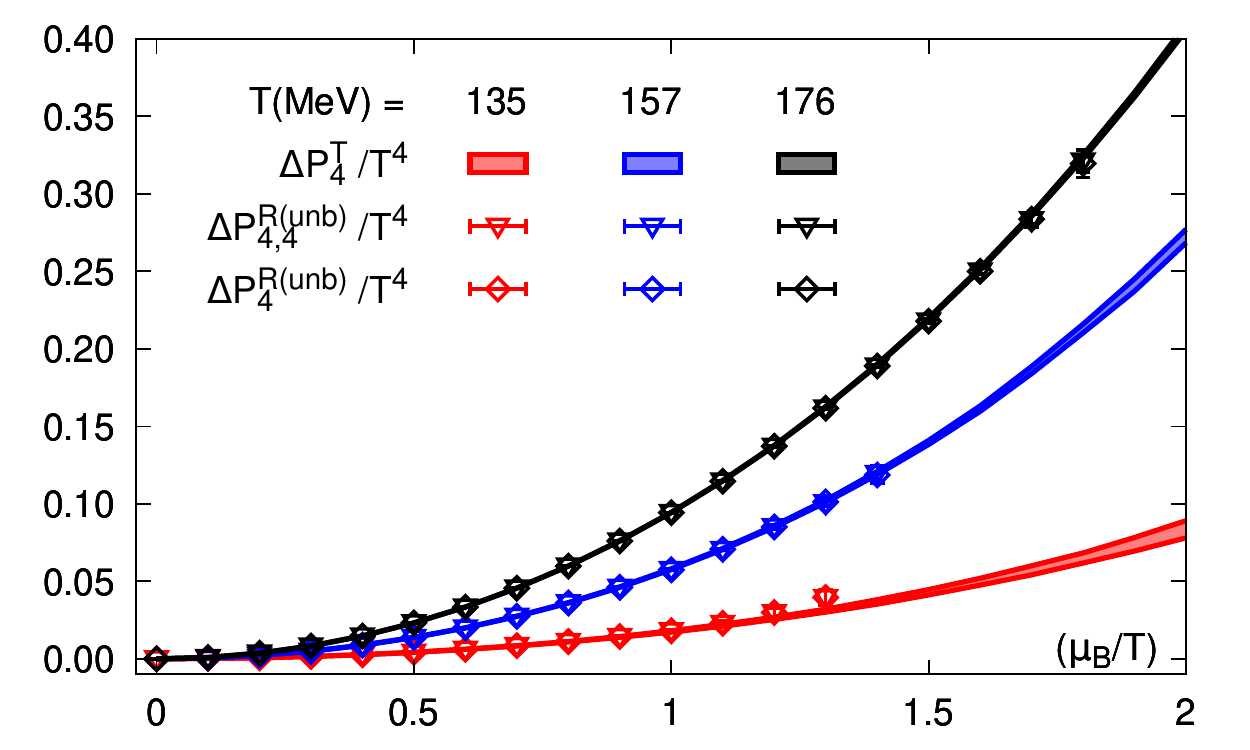}%
\hspace{0.05\textwidth}%
\includegraphics[width=0.40\textwidth]{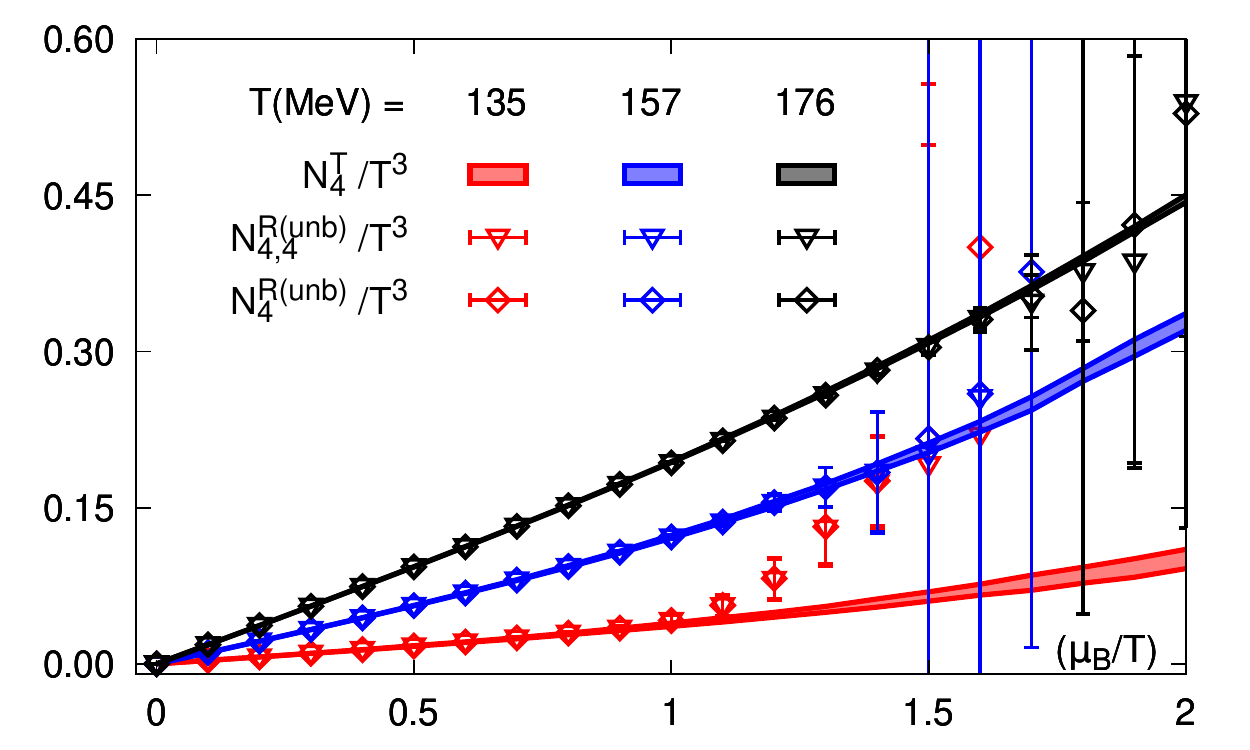}
\caption{$\dP(T,\mB)/T^4$ and $\N(T,\mB)/T^3$ calculated at fourth order in $\mB$ for all the three working temperatures $T=135$, $157$ and $176$ MeV presented in red, blue and black colors respectively.\label{fig:allT}}
\end{minipage}
\end{figure*}

In Fig.~\ref{fig:allT}, we plot the fourth order Taylor expansion and unbiased resummation results for $\dP/T^4$ and $\N/T^3$ as functions of $\hmB\equiv\mB/T$ for all three temperatures viz. $T=135$, $157$ and $176$ MeV. The unbiased resummation results agree quite well with the Taylor series results for $\hmB\lesssim1.1$ - $1.2$. As $\hmB$ is increased however, the resummation calculation breaks down at a value $\hmB=\hmB^c$ that depends upon the temperature. The breakdown happens because the fluctuations of the phase factor $\cos\Theta(T,\mB)$ of the exponential increase rapidly, both in magnitude and sign, as $\hmB$ approaches $\hmB^c$. The increase in fluctuations manifests as a sudden increase in the error bars in the case of the number density, while the pressure becomes indeterminate as the argument of the logarithm (Eq.~\eqref{eq:excess_pressure}) can become negative during bootstrap resampling.

\begin{figure*}[!bht]
\begin{minipage}[!bht]{0.99\textwidth}
\centering
\includegraphics[width=.32\linewidth]{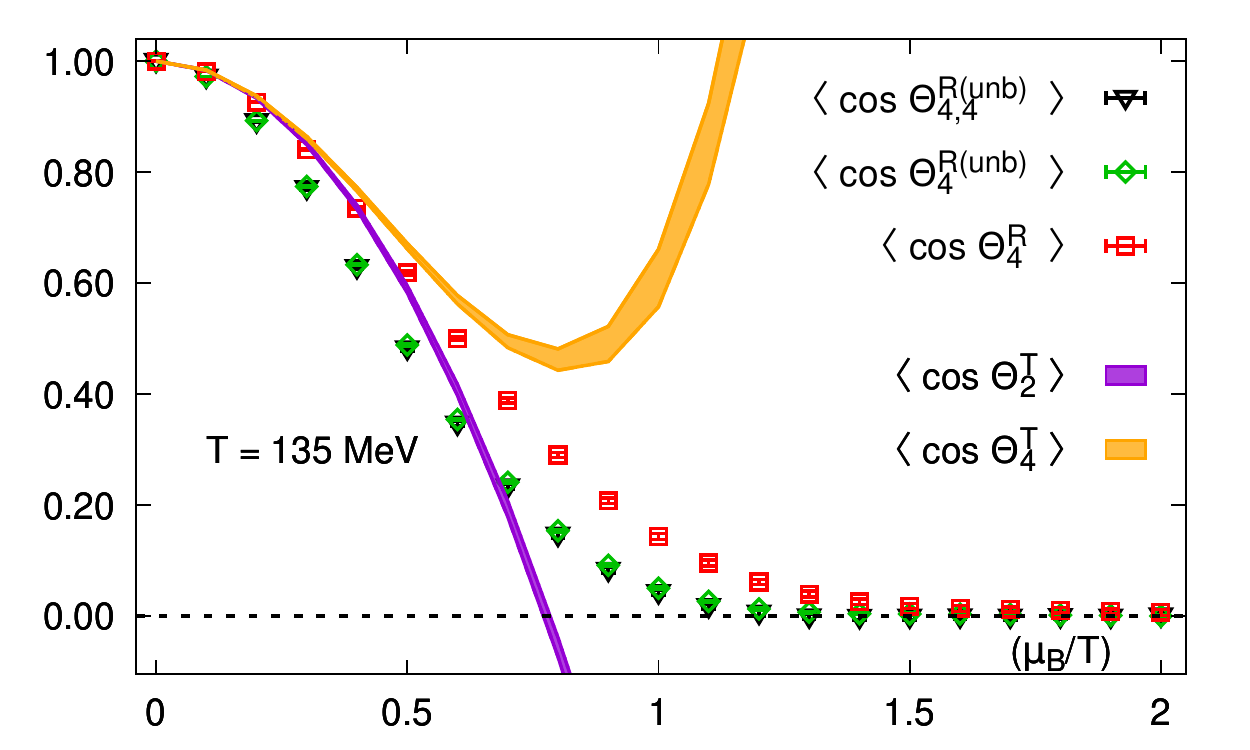}%
\hfill
\includegraphics[width=.32\linewidth]{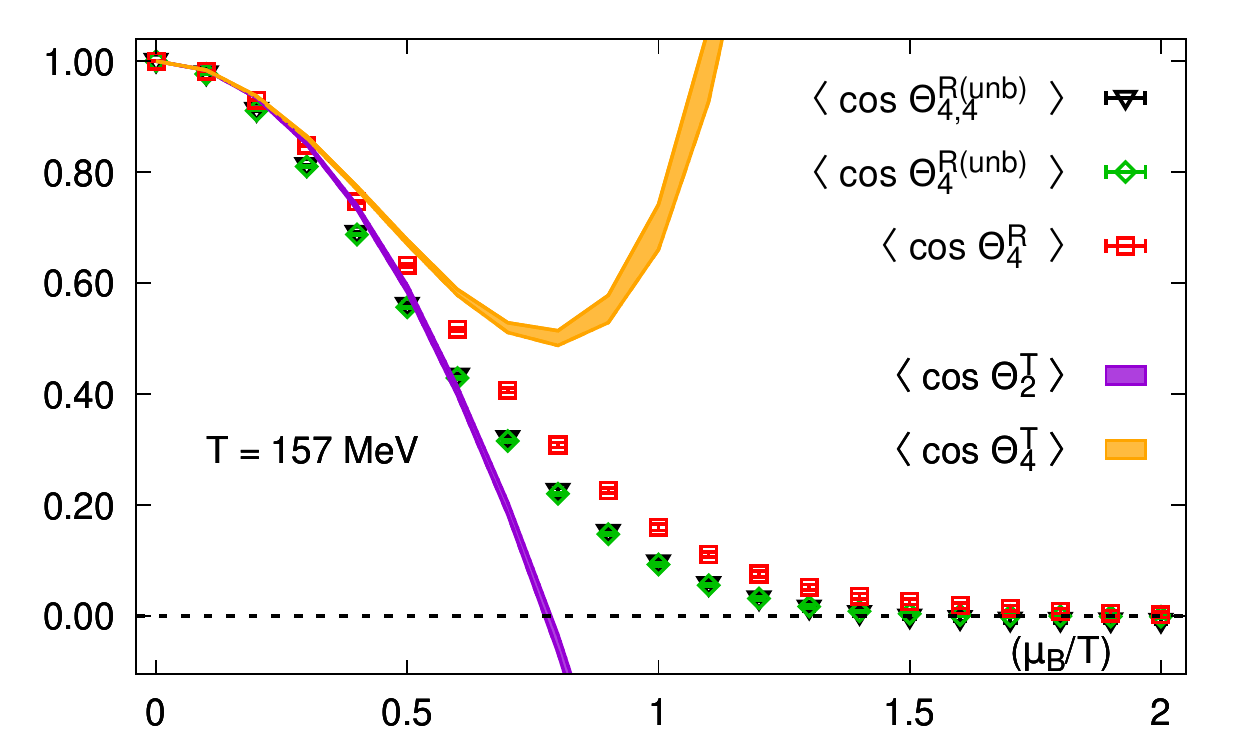}%
\hfill 
\includegraphics[width=.32\linewidth]{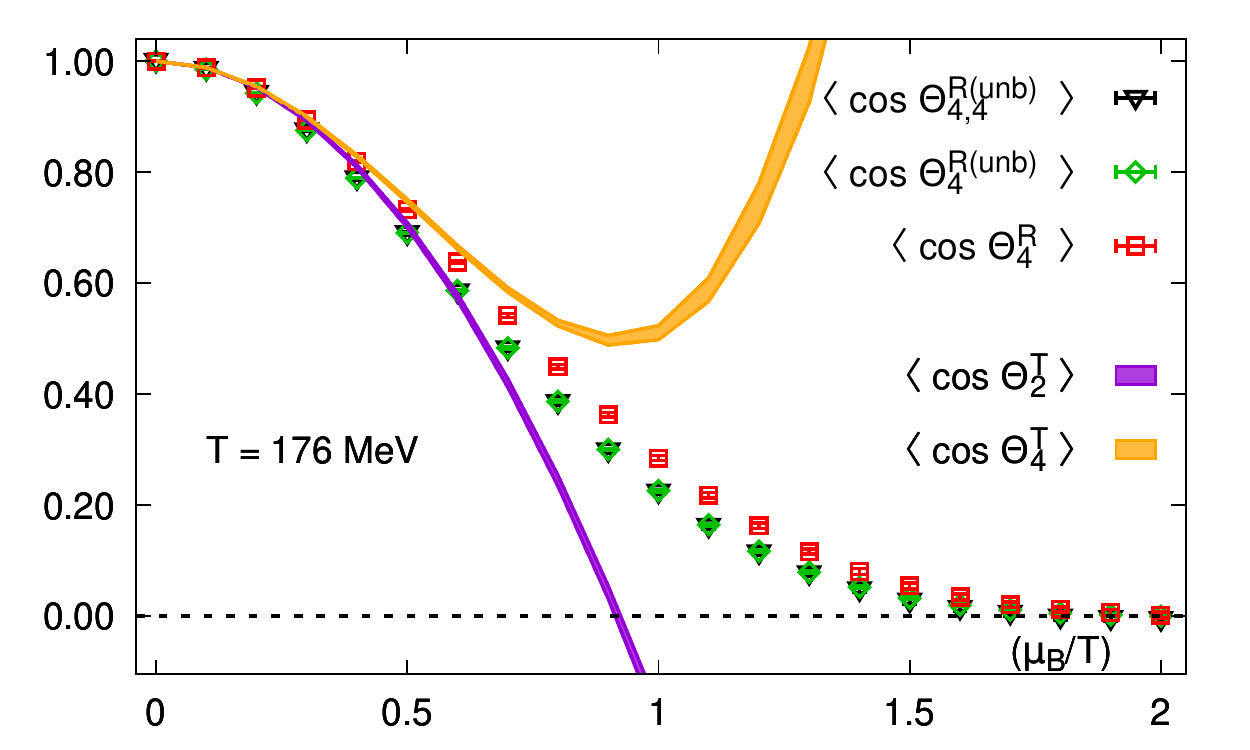}
\caption{Average phasefactor $\left \langle \cos{\Theta}(T,\mB) \right \rangle$ calculated according to Eq.~\eqref{eq:phase_factor} with $N=4$ for $T=135$, 157 and 176 MeV. The second and fourth order Taylor expansion results of $\left \langle \cos{\Theta}(T,\mB) \right \rangle$ are shown as purple and orange bands respectively.}
\label{fig:phase_factor}
\end{minipage}
\end{figure*}

Owing to this increase in fluctuations, the expectation value $\langle\cos\Theta(T,\mB)\rangle$ of the phase factor vanishes for all $\hmB\geqslant\hmB^c$. In Fig.~\ref{fig:phase_factor}, we plot our fourth order results for $\langle\cos\Theta(T,\mB)\rangle$, obtained using Eqs.~\eqref{eq:phase_factor_biased}, \eqref{eq:phase_factor_mu_basis} and \eqref{eq:phase_factor_cumulant_basis} with $N=4$, as a function of $\hmB$ for all three temperatures. We find differences in the biased and unbiased calculations that increase with decreasing temperature and result in different values for $\hmB^c$. Especially at $T=135$~MeV, we see that the unbiased results go to zero around $\hmB\sim1.2$, while the biased results vanish later, around $\hmB\sim1.5$. These differences are important because both the breakdown of the calculation and the vanishing of $\langle\cos\Theta(T,\mB)\rangle$ are expected to occur as $\hmB\to\vert\hat{\mu}_{B0}\vert$, where $\hat{\mu}_{B0}$ is the zero of $\Z(T,\mB)$ that is closest to the origin in the complex $\hmB$ plane~\cite{Mondal:2021jxk}. The origin of the breakdown is thus physical and not a drawback of exponential resummation compared to reweighting or Taylor expansion~\cite{Borsanyi:2022soo}. In fact, since exponential resummation resums the same operators that appear in the Taylor series calculation, the same breakdown should also show up in a Taylor series expansion carried out to sufficiently high order e.g. as a lack of convergence of the Taylor series beyond a certain value of $\hmB$.

\section{Discussion and Outlook}
\label{sec:discussion_and_outlook}
Exponential resummation has been previously introduced as a new way of resumming the finite-density QCD Taylor series~\cite{Mondal:2021jxk}. The contribution of the the $n$th $\hmY$ derivative $D^Y_n(T)$ of $\ln\det M(T,\mY)$, where $M(T,\mY)$ is the fermion matrix, to all orders in $\mY$ is equal to $\exp(D^Y_n(T)\hmY^n/n!)$. In this way, the contribution of the first $N$ derivatives $D^Y_1,\dots,D_N^Y$ that are calculated during the $N$th order Taylor series calculation can be obtained to all orders in $\hmY$. However as the $D^Y_n$ are calculated stochastically, the exponential contains stochastic bias which needs to be subtracted before genuine higher-order contributions can be identified. 

In this paper, we presented a new way of carrying out the exponential resummation in which the stochastic bias was subtracted, at the level of each individual gauge configuration, up to a finite order in $\hmY$ or the cumulant expansion. We applied our formalism to calculate the excess pressure and number density at finite isospin as well as finite baryon chemical potential. Our results were in good agreement with the Taylor series results, both for real as well as imaginary chemical potentials, up to $|\hmI^2|\leqslant4$ (up to $|\hmB^2|\leqslant2$). We also calculated the average phase factor as a function of $\hmB$ using both biased and unbiased resummations. As observed previously~\cite{Mondal:2021jxk}, the vanishing of the phase factor is accompanied by a breakdown of the calculation. The value $\hmB=\hmB^c$ at which the breakdown occurs differs between the biased and unbiased resummations, with the differences increasing as the temperature is decreased. The breakdown of the calculation has physical significance as $\hmB^c$ is the distance from the origin to the closest zero $\hat{\mu}_{B0}$ of the QCD partition function $\Z(T,\mB)$ in the complex $\hmB$ plane. Hence the vanishing of the phase factor could provide yet another way of locating the zeros of $\Z(T,\mB)$ (equivalently, the singularities of $\ln\Z(T,\mB)$ in the complex $\hmB$ plane. Then it would be important to obtain an unbiased determination of $\hmB^c$, especially as the biased and unbiased estimates differ significantly at lower temperatures. 

We also note that with exponential resummation, it is possible to calculate the QCD partition function $\Z(T,\mB)$ itself. By comparison, the QCD Taylor series is an expansion of $\ln\Z(T,\mB)$. The finite Taylor series is analytic over the entire complex $\hmB$ plane, whereas our resummation makes it possible to calculate the zeros of $\Z(T,\mB)$ and hence directly determine the location of poles or branch singularities that could correspond to the location of the much sought after QCD critical point~\cite{Stephanov:2006dn,Fukushima:2010bq,Mukherjee:2019eou}. This has been done previously~\cite{Mondal:2021jxk,Mukherjee:2021tyg}, but we hope to repeat these calculations in the future using our new formalism in order to obtain more reliable estimates of these important observables.


\acknowledgments
We thank the members of the HotQCD collaboration for helpful discussions and for the permission to use their data from the Taylor expansion calculations. The computations in this work were performed using the GPU cluster at Bielefeld University, Germany. We thank the Bielefeld HPC.NRW team for their help and support.

\bibliographystyle{apsrev4-2.bst}

\onecolumngrid
\appendix
\section{Proof of the Unbiasedness of Eq.~\eqref{eq:unbiased_resummed-1} to $\O(\mY^4)$}
\label{app:proof}
In Sec.~\ref{sec:formalism}, we stated without proof that Eqs.~\eqref{eq:unbiased_resummed-1} and \eqref{eq:Cn} (with $N=4$) resum the first four derivatives $D^Y_1,\dots,D^Y_4$ in such a way that the resulting exponential as well as the excess pressure are both unbiased to $\O\left(\mY^4\right)$ where $Y \equiv B,I$. To see why this is so, we start by Taylor-expanding the exponential in Eq.~\eqref{eq:unbiased_resummed-1}. To $\O\left(\mY^4\right)$, one obtains (with $\hmY \equiv \mY/T)$:
\vspace{.2cm}
\begin{align}
\exp\left[\sum_{n=1}^4\frac{\C^Y_n(T)}{n!}\mT^n\right] =\;\sum_{k=0}^\infty\,\frac{1}{k!}\,\left[\sum_{n=1}^4\frac{\C^Y_n(T)}{n!}\mT^n\right]^k = 1 + \sum_{k=1}^4 \, \A^Y_k(T)\,\frac{\hmY^{\,k}}{k!} + \O\left(\hmY^5\right), \\ \notag
\end{align}
where the $\A^Y_k$, $k=1,\dots,4$ are given by
\begin{align}
\notag \\
    \A^Y_1(T) &\;=\; \overline{D^Y_1},\notag \\
    \A^Y_2(T) &\;=\; \overline{D^Y_2} + \overline{(D^Y_1)^2}, \notag \\
    \A^Y_3(T) &\;=\; \overline{D^Y_3} + 3\,\overline{D^Y_2 D^Y_1} + \overline{(D^Y_1)^3}, \notag \\
    \A^Y_4(T) &\;=\; \overline{D^Y_4} + 3\,\overline{(D^Y_2)^2} + 4\,\overline{D^Y_3 D^Y_1} + 6\,\overline{D^Y_2 (D^Y_1)^2} + \overline{(D^Y_1)^4}.
\end{align}
\vspace{.4cm}
We note that the $\A^Y_k$ are just the derivatives of $\det\M$ w.r.t. $\hmY$~\cite{Allton:2005gk} 
\begin{align}
    \A^Y_k(T) \equiv \frac{\partial^k}{\partial \hmY^k} \Big[\det \M(T,\mY)\Big]_{\mY=0}, \\ \notag
\end{align}
but with the terms appearing in the derivative evaluated in an unbiased manner. Now, as per Eq.~\eqref{eq:unbiased_resummed-1}, we need to extract the real part of the exponential. This means that the above series becomes an even series in $\hmY$, since the coefficients of even (odd) powers of $\mY$ are purely real (imaginary). We therefore have: 
\begin{equation}
\frac{\Delta P_{4}^{R(\text{unb})}}{T^4} = \frac{\Nt^3}{\Ns^3}\,\ln\left\langle 1 + \sum_{k=1}^2 \, \A^Y_{2k}(T)\,\frac{\hmY^{\,2k}}{(2k)!} + \O\left(\hmY^6\right) \right\rangle.
\label{eq:PT4ln}
\end{equation}
We compute $\Delta P_{4}^{R\text{(unb)}(T,\mY)}/T^4$ in the above equation by using the well-known formula for $\ln(1+x)$, namely
\begin{equation}
    \ln (1+x) = x - \frac{x^2}{2} + \Ob(x^3).
    \label{eq:logarithm_formula}
\end{equation}
Collecting coefficients upto $\Ob(\mY^4)$, we find the following:
\begin{equation}
    \frac{\Delta P_{4}^{R(\text{unb})}}{T^4} = \frac{\Nt^3}{\Ns^3} \left[
    \frac{\langle\A^Y_2\rangle}{2!} + \frac{\langle\A^Y_4\rangle - 3\langle\A^Y_2\rangle^2}{4!}\right] + \O(\mY^6).
\label{eq:produce_Taylor_coefficients}
\end{equation}
This is just the Taylor series expansion $\dP^T_4(T,\mB)$ of the excess pressure to fourth order i.e.
\begin{equation}
    \frac{\Delta P_{4}^{R(\text{unb})}}{T^4} = \frac{\chi_2^Y(T)}{2!}\mT^2 + \frac{\chi^Y_4(T)}{4!}\mT^4 + \O(\mY^6),
\end{equation}
with the Taylor coefficients given by the usual formulas~\cite{Allton:2005gk}
\begin{equation}
    \chi_2^Y = \frac{\Nt^3}{\Ns^3}\,\big\langle\A^Y_2\big\rangle \quad \text{and} \quad
    \chi_4^Y = \frac{\Nt^3}{\Ns^3} \bigg(\big\langle\A^Y_4\big\rangle - 3\big\langle\A^Y_2\big\rangle^2\bigg).
\end{equation}
Thus we find that Eq.~\eqref{eq:unbiased_resummed-1} reproduces the usual Taylor series expansion of the excess pressure to $\O\left(\mY^4\right)$. Since the Taylor coefficients are calculated in an unbiased manner, we conclude that the exponential in Eq.~\eqref{eq:unbiased_resummed-1} too is unbiased to $\O\left(\mY^4\right)$.
\end{document}